\documentclass[12pt]{article}  
\usepackage[dvips]{epsfig}
\usepackage{amssymb}
\usepackage{cite}
\newcommand{\vev}[1]{\left\langle #1\right\rangle}

\newcommand{\GeV}{\; \mathrm{GeV}} 
 
\newcommand{\lapproxeq}{\lower .7ex\hbox{$\;\stackrel{\textstyle  
<}{\sim}\;$}} 
\newcommand{\gapproxeq}{\lower .7ex\hbox{$\;\stackrel{\textstyle  
>}{\sim}\;$}} 
\newcommand{\stackdown}[2]{\lower 1.4ex\hbox{$\;\stackrel{\textstyle{#1}}  
{\scriptstyle{#2}}\;$}}
\newcommand{\beq}{\begin{equation}} 
\newcommand{\eeq}{\end{equation}} 
\newcommand{\bea}{\begin{eqnarray}} 
\newcommand{\eea}{\end{eqnarray}}

\newcommand{\etal}{\textit{et. al.}}
\newcommand{\cosb}{\cos\beta}
\newcommand{\sinb}{\sin\beta}

\newcommand{\mneu}{ m^2_{\tilde{Z}_a} }

\begin{document} 
\begin{titlepage} 
 
%%%%%%%%%%% 
%%%%%%%%%%% 
\begin{flushright} 
\parbox{6.6cm}{ ACT-16/00, CTP-TAMU-36/00 \\ UA/NPPS-BSM-1-00 \\
hep-ph/0011370 } 
\end{flushright} 
%%%%%%%%%% 
%%%%%%%%%%
\begin{centering} 
\vspace*{1.5cm} 
 
{\large{\textbf {On the Radiative Corrections to the 
Pseudo-scalar Higgs Boson Mass}}}\\
\vspace{1.4cm}

{\bf A.~Katsikatsou} $^{1}$,
{\bf A.~B.~Lahanas} $^{1}$, \, 
{\bf D.~V.~Nanopoulos} $^{2}$  \, and \, {\bf V.~C.~Spanos} $^{1}$  \\ 
\vspace{.8cm} 
$^{1}$ {\it University of Athens, Physics Department,  
Nuclear and Particle Physics Section,\\  
GR--15771  Athens, Greece}\\ 
 
\vspace{.5cm} 
$^{2}$ {\it Department of Physics,  
         Texas A \& M University, College Station,  
         TX~77843-4242, USA, 
         Astroparticle Physics Group, Houston 
         Advanced Research Center (HARC), Mitchell Campus, 
         Woodlands, TX 77381, USA, and \\ 
         Academy of Athens,  
         Chair of Theoretical Physics,  
         Division of Natural Sciences, 28~Panepistimiou Avenue,  
         Athens 10679, Greece}  \\ 

\end{centering}
\vspace{3.cm} 
%%%%%%%%%%%%%%%%%%%%%%%%% 
\begin{abstract} 
We reexamine the one-loop corrections to the mass of the 
pseudo-scalar Higgs boson, using the effective potential. In the absence
of the chargino and neutralino contributions 
its mass exhibits a large scale dependence in the large
$M_{1/2}$ regime, especially  for values of $\tan \beta>20$.
Thus, although of electroweak origin, the heaviness
of the $M_{1/2}$, in conjunction with the largeness of $\tan \beta$, 
makes these corrections very important for establishing a scale
independent result and an unambiguous determination of the pseudo-scalar
mass in this region of the parameter space. 
\end{abstract} 
\end{titlepage} 
\newpage 
\baselineskip=18pt
%%%%%%%%%%%%%%%%%%%%%%%%%% paper body %%%%%%%%%%%%%%%%%%%%%%%%
%%%%%%%%%%%%%%%%%%%%%%%%% section 0 %%%%%%%%%%%%%%%%%%
\section{Introduction}

The Higgs sector of the MSSM has been put under experimental scrutiny
the last few years and reliable lower bounds on the Higgs masses 
have been already established imposing tight constraints on supersymmetric
models \cite{bounds}.

Radiative corrections to the Higgs masses have been extensively studied, 
especially for the lightest $CP$-even Higgs state
for which a theoretical upper
bound has been established of $\simeq 130 \GeV$
\cite{Haber,Ellis1,Barbieri,Eberl,Brignole,Drees,Pokorski,hoang,casas,%%
Bagger,Pierce,zhang,hhw,espinosa}.
Among the others states the
$CP$-odd Higgs $A$ may play a significant role in processes in which 
$CP$ is violated \cite{Pilaftsis}, and also in 
cosmology since the lightest supersymmetric particle (LSP)
pair annihilation to a fermion pair through $A$ exchange
may enhance the corresponding cross section for large $tan \beta$, yielding
LSP relic densities compatible with the recent astrophysical data
\cite{spanos}.
The latter process
is sensitive to the mass of the pseudo-scalar Higgs and a reliable
determination of its mass is highly demanded.

Existing studies in literature discuss the radiative corrections to
the pseudo-scalar Higgs mass using either the effective potential approach
or by studying the loop corrections to the propagator
\cite{Haber,Ellis1,Barbieri,Brignole,Drees,Pokorski,Bagger,Pierce,Eberl}.
The mass in the second approach is the physical, or pole, mass
$m_A (pole)$ which is certainly
scale independent and differs from that calculated using the effective
potential denoted by $m_A$. The two masses are related
by  \cite{Ellis1,Brignole}
\bea
m_A^2 (pole) = m_A^2 +
{{\Pi}_{AA}}(0) - {{\Pi}_{AA}}(m_A(pole)) \;,    \label{prwth}
\eea
where $ {\Pi}_{AA}(p^2) $ are the corrections to the pseudo-scalar
propagator at momentum $p$.
In some studies it is claimed that $m_A$ is scale independent, however
this assertion is not entirely correct since the difference
${{\Pi}_{AA}}(m_A(pole)) -  {{\Pi}_{AA}}(0) $ includes small logarithmic
parts, which depend explicitly on the scale $Q$.
For completeness these should be added to $ m_A$ to render scale invariance.

In this note we consider all radiative corrections to the pseudo-scalar
mass as derivatives of the one-loop effective potential. We particularly
show the importance of the chargino and neutralino corrections in
establishing a result which is scale independent and approximates accurately 
the pole mass. Long known third
generation sfermion contributions are not by themselves
adequate to yield a scale independent result for large $M_{1/2}$ values.
The reason is that the Renormalization Group Equation 
(RGE) of the Higgs mixing parameter $m_3^2$, entering into the expression
for $m_A$, includes terms
%%%%%%%%%%%
\bea
{\frac{1}{16 \pi^2}}\; \mu\; (-6 g_2^2 M_2 -{\frac{6}{5}} g_1^2 M_1) \,,
\label{rge}
\eea
%%%%%%%%%
resulting to a scale dependence of $m_3^2$ 
not cancelled by the third generation
sfermion corrections alone. In the large  $M_{1/2}$ and large $\tan \beta$
regime such corrections are not numerically small.

In our approach for the determination of the
pseudo-scalar mass we have  duly taken into account all contributions,  
including the chargino/neutralino corrections,
and have observed that the latter contribute significantly to the
stabilization of the pseudo-scalar Higgs boson  mass with respect the
scale $Q$.
If these contributions were neglected stabilization would be spoiled
in the large $M_{1/2}$ region, where the running of the parameter 
${m_{3}^{2}}(Q)$ due to the gauginos becomes important. The situation
would be even more dramatic if in addition to having large $M_{1/2}$  we are
in the large $\tan \beta $ regime where such deviations from
stability are enhanced as being proportional to $\tan \beta $.

From Eq. (\ref{prwth}) we see that the effective potential mass
differs from the physical mass by $\Pi (0) - \Pi (m_A^2)$.
In our scheme, and for a more reliable estimate of the pseudo-scalar 
mass, besides the chargino and neutralino corrections we have also included
the contributions of the remaining sectors to $m_A$ as well as the leading
logarithmic corrections 
of $\Pi (0) - \Pi (m_A^2)$. 
Including these the resulting mass is certainly scale independent, 
to this loop order, and is found to be very close
to the one-loop pole mass. Differences are less than $2 \%$.
In this way we have approximated satisfactorily
the pole mass and avoided the complexities of 
calculating one-loop integrals, which are usually
expressed by the Passarino--Veltman functions.

%%%%%%%%%%%%%%%%%%%%%%%%%%%%%%%%%%%%%%%%%%%%%%%%%%%%%%

%%%%%%%%%%%%%%%%%%%%%%%%%%%%%%%%%%%%%%%%%%%%%%%%%%%%%%
\section{Radiative Corrections to the pseudo-scalar Higgs Mass}
We assume a low energy supersymmetric theory described by the superpotential 
\bea
{\cal W} = h_t H_2^T  \epsilon \,Q U^c +
               h_b H_1^T  \epsilon \,Q D^c +
               h_\tau  H_1^T  \epsilon \,L E^c +
               \mu  H_2^T  \epsilon \,H_1 \, ,
\eea
where the elements of the antisymmetric $2 \times 2$ matrix $\epsilon$ are
given by $ \epsilon_{12}=-\epsilon_{21}=-1$. In it 
only the dominant Yukawa terms of the third generation are assumed
nonvanishing.  \\
The soft-SUSY breaking part of the Lagrangian is given by, in an obvious
notation,
%%%%%%%%%%%%%%%%%
\bea
{\cal L}_{\mathrm{scalar}} = &-& {\sum_i}\;m_{\phi_i}^2 |{\phi_i}|^2
\;-\; ( m_3^2 H_2^T  \epsilon \,H_1 + \,h.c)
\nonumber \\
&-&\,(A_t h_t H_2^T  \epsilon \,Q U^c +
    A_b h_b H_1^T  \epsilon \,Q D^c +
    A_\tau h_\tau H_1^T  \epsilon \,L E^c + \, h.c) \nonumber \\ 
&-& {\frac{1}{2}}\;( M_1{\tilde B} {\tilde B} +
M_2 {\tilde W}^{(i)} {\tilde W}^{(i)} + M_3 {\tilde G} {\tilde G}
+\, h.c.)  \,. \nonumber 
\eea
%%%%%%%%%%%%%%%%%%%%%%%%%

At the tree level the scalar potential of the theory is given by 
\bea
V^{0} & = & m_1^2 \, |H_1^0|^2 + m_2^2 \, |H_2^0|^2 
           + m_3^2 \, (H_1^0 H_2^0 +h.c) \nonumber\\
  & + & ({{g^2+g'^2}\over{8}})(|H_1^0|^2 - |H_2^0|^2)^2\, ,
\eea
where $m_i^2 \equiv m_{H_i}^2+\mu^2$. In this equation we have only kept 
the part that depends exclusively on the neutral components of the
Higgs fields $H_{1,2}^0$ which is relevant for the electroweak
symmetry breaking.

The one-loop correction to the effective potential
is 
\beq 
 \Delta V^1 = {1\over{64\pi^2}} \,
\sum_J (-1)^{2s_J} \, (2s_J+1)\, m^4_J \,\left( \ln {m^2_J\over {Q^2}} 
- {3\over {2}} \right) \, , \label{dv1}
\eeq
where $\;m_J\;$ are field dependent masses and $s_J$ we denotes the
spin of the $J$-particle. 

The minimization  of the one-loop corrected
effective potential $V^1 \equiv V^0 + \Delta V^1$ yields the following
relations 
\beq
\sin 2\beta = -{ {2 m_3^2}\over{\bar{m}_1^2+\bar{m}_2^2} } \,\,, \,\,\,
v^2={ 8\over{g^2+g'^2} } { {\bar{m}_1^2 - \bar{m}_2^2 \tan^2\beta} \over
 {\tan^2\beta-1} } \, , \label{min}
\eeq
where we have defined
%%%%%%%%%%%%%%%%%%%%%%%%%%%%%%%%%%%%%%%%%
\footnote{Note that in our notation 
$v_i \equiv \vev{H^0_i}$, $v_1 \equiv \frac{v}{\sqrt{2}} \cos\beta$,
$v_2 \equiv \frac{v}{\sqrt{2}}\sin\beta$,
$M_W^2=g^2(v_1^2+v_2^2)/2=g^2v^2/4$ .}
%, $v^2 \equiv v_1^2+v_2^2$,
%$M_Z^2={ {g^2v^2} \over{2\cos^2\theta_w} }$,
%$g^2+g'^2 = { {g^2} \over{\cos^2\theta_w} }$. }
%%%%%%%%%%%%%%%%%%%%%%%%%%%%%%%%%%%%%%%%%%%%%%%%%%%%%%%
\beq
\bar{m}_i^2 \equiv m_i^2 + \Sigma_i \,\, , \,\,\,
\Sigma_i \equiv \left. { {\partial V^1}\over{\partial (\mathrm{Re} H_i^0)^2 }} 
\right|_{\vev{H_i^0}} \,. \label{sigma}
\eeq

As discussed in the introduction the relation between the pole and the mass
calculated using the effective potential is given by Eq. (\ref{prwth}).
In this $ m_A^2 $ is the non-zero eigenvalue of the matrix
$
{1\over{2}} \left. { {\partial^2 V^0}\over{\partial \phi_i \partial \phi_j} }
\right|_{\vev{H_i^0}} 
$
where $\phi_i \equiv \mathrm{Im}H_i^0$.  
At the tree level 
\beq
{1\over{2}} \left. { {\partial^2 V^0}\over{\partial \phi_i \partial \phi_j} }
\right|_{\vev{H_i^0}} =
(m_1^2+m_2^2) \; \cosb\sinb
               \left( \begin{array}{cc} 
         \tan\beta & 1\\
          1 & \cot\beta  
               \end{array} \right) \, . \label{matree}
\eeq
%%%%%%%%%%%%%%%%
The matrix above has one zero eigenvalue corresponding to the mass of the 
Goldstone mode which is eaten up by the $Z$-boson. 
The other eigenvalue is the mass squared of the pseudo-scalar Higgs,
given by $m_A^2=m_1^2+m_2^2$ in the lowest order.

At the one-loop the Eq.~(\ref{matree}) is modified to
\bea
{1\over{2}} \left. { {\partial^2 V^1}\over{\partial \phi_i \partial \phi_j} }
\right|_{\vev{H_i^0}} &=&
(\bar{m}_1^2+\bar{m}_2^2) \; \cosb \sinb
              \left( \begin{array}{cc} 
         \tan\beta & 1\\
          1 & \cot\beta  
               \end{array} \right) \nonumber\\ 
 &+& {1\over{64\pi^2}} \,
\sum_J (-1)^{2s_J} \, (2s_J+1)\, m^2_J \,\left( \ln {m^2_J\over {Q^2}} 
- 1 \right) \nonumber\\ 
&\times& \left( \begin{array}{cc} 
  { {\partial^2 m^2_J}\over{\partial\phi_1^2} }
 -2 { {\partial m^2_J}\over{\partial (\mathrm{Re} H_1^0)^2 }} & 
       { {\partial^2 m^2_J}\over{\partial\phi_1\partial\phi_2} } \\
  { {\partial^2 m^2_J}\over{\partial\phi_2\partial\phi_1} } & 
 { {\partial^2  m^2_J}\over{\partial\phi_2^2} }
 -2 { {\partial m^2_J}\over{\partial (\mathrm{Re} H_2^0)^2 }}
       \end{array} \right)_ {\vev{H_i^0}}\, , \label{master}
\eea
where we have used the one-loop minimization conditions of 
Eq.~(\ref{min}) as well as the relations of Eq.~(\ref{sigma})
and the one-loop corrections to the scalar potential given by the 
Eq.~(\ref{dv1}). 
This is the master formula we are going to use for
the calculation of the one-loop corrections to the
$CP$-odd Higgs boson in the effective potential approach.

Using the minimization condition relating $(\bar{m}_1^2+\bar{m}_2^2) $ to
the Higgs mixing parameter $ {m}_3^2 $ (see Eq. (\ref{min})), the above
formula can always be cast into the form
%%%%%%%%%%%%%%%%
\bea
{1\over{2}} \left. { {\partial^2 V^1}\over{\partial \phi_i \partial \phi_j} }
\right|_{\vev{H_i^0}} &=&
-({m}_3^2+\Delta) 
              \left( \begin{array}{cc} 
         \tan\beta & 1 \\
          1 & \cot\beta  
               \end{array} \right)  , \label{delta}
\eea
%%%%%%%%%%%%%
resulting to a pseudo-scalar mass squared $m_A^2$ given by
%%%%%%%%%%%%%5
\bea
m_A^2 \;=\; - \frac{ 2 ( m_3^2 + \Delta )}{ \sin 2 \beta } \; . \label{mass}
\eea
%%%%%%%%%
For the calculation of the $CP$-odd Higgs mass we thus need calculate the
contribution of each particle species to $\Delta$. The dominant
third generation sfermion 
contributions to this quantity have been long known. The neutralino and
chargino contributions are claimed to be small and hence not expected to
yield substantial corrections to the pseudo-scalar mass \cite{Eberl}.
However this may not
be true when the soft gaugino mass $M_{1/2}$ is large as pointed out in the
introduction. In this regime such
corrections may be sizeable and should be duly taken into account.

It is well known that 
radiative corrections to $m_A$, calculated through the
effective potential, are not stable with changing the renormalization scale
(see for instance Ref. \cite{baer} and references therein).
Empirically, one calculates $m_A$  at an average stop mass scale, 
$Q_{\tilde t} \approx \sqrt{m_{{\tilde t}_{1}}\;m_{{\tilde t}_{2}}}$,
in which case the radiative corrections of the
third generation sfemions are small and can be safely neglected.
In this case the pseudo-scalar mass squared is given by
$m_{A}^{2} =-2 \, {{m_{3}^{2}}(Q_{\tilde t})}/{{\sin 2 \beta} (Q_{\tilde t}) } $, where
$m_{3}^{2} $ is the Higgs mixing parameter. 
 Although in principle this is correct, the contributions of charginos and
neutralinos are not small at this scale, especially 
when $M_{1/2}$ is large. In some cases this
may produce an error in the determination of its mass as large as
$25 \%$.
Excluding the chargino/neutralino contribution is legitimate
provided the relevant scale is not taken to be $Q_{\tilde t}$, but rather
the average of the chargino and neutralino masses $Q_{\tilde \chi}$ 
defined by the following expression
%%%%%%%%%%%
\bea
Q_{\tilde \chi}^2\;=\; \frac{1}{2}\;( \;\vev{m_{\tilde C}^2}\;+\;
\vev{m_{\tilde Z}^2} \;)  \; .
%%Q_{\tilde \chi}^2\;=\; \frac{1}{6} \; (\;\sum_{i}\; m_i^2\;)   \;.
\eea
%%%%%%%%%%%
In the definition above $ \vev{m_{\tilde C}^2} , \vev{m_{\tilde Z}^2}\;$
denote the averages of the chargino and neutralino masses squared.  
At this scale it is legitimate to neglect their 
contributions. This scale however may differ
substantially from $Q_{\tilde t}$, when $M_{1/2}$ is large. 
Therefore one expects large variations of the $\;m_A\;$ between these
two scales, if only the third generation sfermion contribution is kept,
resulting to a poor determination of the pseudo-scalar mass.
In the
following we shall calculate all corrections to the pseudo-scalar mass.
The gauge and Higgs boson contributions to $\Delta$ although less important
should be also included to yield a result that is scale independent and
approximates satisfactorily well the pole mass.

An alternative way to calculate the quantity 
$\Delta$ is through the one-loop
corrections to the pseudo-scalar propagator $\Pi_{AA}$. In fact following
Ref. \cite{Bagger} we find that $\Delta$ can be expressed as
\bea
\Delta\;=\; s_{\beta}\;c_{\beta}\;(\;  \Pi_{AA}(0) \;-\;
s^2_{\beta}\;\frac{t_1}{v_1}\;-\;c^2_{\beta}\;\frac{t_2}{v_2}\;)
\label{alter}
\eea
where $\;s_{\beta} \equiv \sin \beta\;,\; c_{\beta} \equiv \cos \beta\;$,
 and 
$\;t_{1,2}\;$ are the one-loop contributions to the v.e.v's
$\;\vev{H_{1,2}}\;$. The latter can be calculated by either using the
effective potential \cite{nath}
or by employing diagrammatic techniques. In principle the
two approaches yield identical results.
The gauge dependence of the tadpole graphs cancels against gauge dependence
of loop diagrams in $\Pi_{AA}$ which involve gauge and Goldstone bosons.
Lacking the form of the effective potential in gauges other than the Landau
gauge, such as the popular t' Hooft gauge for instance,  
we prefer to rely on diagrammatic techniques to calculate the gauge and
Higgs contributions to the tadpoles whenever needed.

In the following we shall discuss the various contributions to the
quantity $\Delta$. 
To establish our notation we briefly recall 
the results of Refs.~\cite{Ellis1,Brignole}, and we first consider the third
generation sfermion contributions to the $CP$-odd Higgs scalar. 
The one-loop corrections are given by Eq. (\ref{delta})  with $\Delta$
given by
%%%%%%%%%%%%%%%%%%%%
\bea
\Delta^{\tilde{q}} = {1\over{32\pi^2}} \; \mu\;
\sum_{f=t,b,\tau} 
    {N_f \;h_f^2 \; A_f  \over{ m_{\tilde{f}_1}^2 - m_{\tilde{f}_2}^2 } }
         [f(m_{\tilde{f}_1}^2)-f(m_{\tilde{f}_2}^2)] . \nonumber 
\eea
%%%%%%%%%%%%%
In the equation above $\;N_f\;$ is the color factor and 
the function  $f(m^2)$ is defined by
%%%%%%%
\beq
f(m^2) \equiv 2m^2 \left[ \ln {m^2 \over{Q^2}} -1 \right] \,.  \label{funf}
\eeq
%%%%%%%%%%%%
The remaining sfermion contributions are zero because of the vanishing of 
the corresponding Yukawa couplings. 

For the chargino contribution we need their field dependent mass matrix
$ \mathcal{M}_C\;$. Its squared
$\;\mathcal{M}_C^2=\mathcal{M}_C^\dagger \mathcal{M}_C \;$ 
is given by the following relation
%%%%%%%
$$
\left( \begin{array}{cc}
 M_2^2 + g^2|H_2^0|^2  &  -g( M_2 H_1^0+\mu H_2^{0*}) \\
 -g(M_2 H_1^{0*} + \mu H_2^{0}) & \mu^2 + g^2|H_1^0|^2 
 \end{array} \right) \, . 
$$
%%%%%%%%%%%%
From this, after a straightforward calculation, 
it follows that the chargino contribution to the quantity $\Delta$ is
given by
\beq
\Delta^{\tilde{C}} \equiv - {g^2 \over{16\pi^2}} 
    {  M_2 \mu \over{ m_{\tilde{C}_1}^2 - m_{\tilde{C}_2}^2 } }
         [f(m_{\tilde{C}_1}^2)-f(m_{\tilde{C}_2}^2)] \,.
\eeq
%%%%%%%%%%%%%%%%%%%%%%%%%%%%%%%%%%%%%

%%%%%%%%%%%%%%%%%%%%%%%%%%%%%%%%%%%%%%%%%%%%%%%%%%%%%%%%%% 
 
%\subsection{Neutralino contributions}
The contribution of the neutralinos to $\Delta$ is less trivial
to be found. Their field dependent mass matrix can be put in the following
form
\beq 
\mathcal{M}_N= 
\left( 
\begin{array}{cccc} 
   M_1 & 0 & {g'H_1^{0*}\over{\sqrt{2}}} & -{g'H_2^{0*}\over{\sqrt{2}}} \\ 
   0 & M_2 & -{gH_1^{0*}\over{\sqrt{2}}} & {gH_2^{0*}\over{\sqrt{2}}}   \\ 
   {g'H_1^{0*}\over{\sqrt{2}}} & -{gH_1^{0*}\over{\sqrt{2}}} & 0 & -\mu \\ 
   -{g'H_2^{0*}\over{\sqrt{2}}} & {gH_2^{0*}\over{\sqrt{2}}} & -\mu & 0 
\end{array} 
\right) \,. 
\eeq
whose squared is defined by
\beq 
\mathcal{M}_N^2=\mathcal{M}_N^\dagger \mathcal{M}_N \, . 
\eeq 
Its field dependent eigenvalues $\mneu \;,\;a=1,2,3,4 $ are real and are
determined from the eigenvalue equation
\bea 
h(\lambda) \equiv
  \det(\mathcal{M}_N^2 -\lambda I) = 0 . \label{det} 
\eea 
Obviously $\mneu $ are the physical neutralino masses squared when the
Higgs fields are on their v.e.v's. 
The function $h(\lambda)$ can be written as 
\beq 
h(\lambda)=\lambda^4+A\lambda^3+B\lambda^2+C\lambda+D \, , 
\label{eigen} 
\eeq 
where the coefficients $A$, $B$, $C$ and $D$ depend 
on the neutral Higgs fields $H_i^0$ .

In order to bring the second derivatives of the potential 
into the form of Eq.~(\ref{delta}) we need the  derivatives
$ {\partial \mneu}\over{\partial (\mathrm{Re} H_i^0)^2 } $ and 
${\partial \mneu}\over{\partial \phi_i \partial \phi_j} $
where $\phi_i$ is 
$ \mathrm{Im} H_i^0 $. The relevant derivatives entering the Eq.~(\ref{master}) are 
%%%%%%%
\beq 
\left. { {\partial \mneu}\over{\partial (\mathrm{Re} H_i^0)^2 }}\right|_{\vev{H_i^0}} = 
-{1\over{h'(\mneu)}} 
\left[(\mneu)^3 \dot{A}_i  + (\mneu)^2 \dot{B}_i  + 
(\mneu) \dot{C}_i  +  \dot{D}_i \right] \,, 
\label{real} 
\eeq 
%%%%%%%%%%%
and also 
\beq  \left.
{ {\partial^2 \mneu}
       \over{\partial \phi_i \partial \phi_j} }\right|_{\vev{H_i^0}}=
-{1\over{h'(\mneu)}} 
\left[(\mneu)^3 A''_{ij}+(\mneu)^2 B''_{ij}+(\mneu) C''_{ij}+D''_{ij} \right]
\,. \label{imag} 
\eeq 
%%%%%%%%%%%%%%%%%%%%%%
In these equations 
\beq 
\left. \dot{A}_i \equiv 
{{\partial A}\over{\partial (\mathrm{Re} H_i^0)^2 } }\right|_{\vev{H_i^0}}\,
\; \mathrm{and} \;
\left. A''_{ij} \equiv
{{\partial^2 A}\over{\partial \phi_i \partial \phi_j  } }\right|_{\vev{H_i^0}}
\, ,
\label{ndef1} 
\eeq
and the same holds for $B$, $C$ and $D$. In Eqs. (\ref{real},\ref{imag}) 
$h'(\mneu)$ stands for the derivative
\beq 
h'(\mneu) \equiv \left.{ 
{dh(\lambda)}\over {d\lambda} } \right|_{\lambda=\mneu,\vev{H_i^0}}\;\; . 
\label{ndef11} 
\eeq 
In these all Higgs fields are meant on their v.e.v's and hence all masses
appearing are the physical tree level neutralino masses.
In deriving Eq.~(\ref{imag}) we have used the fact that
\beq 
A'_i=B'_i=C'_i=D'_i=0 \;,\;\;\mbox{when}\;\;\phi_i=0\,. 
\eeq 
Using these one can find that the neutralino contribution to $\;\Delta \;$
is given by
\beq 
\Delta^{\tilde{Z}} \equiv - {g^2\over{32\pi^2}}  
      \,   \mu  \, \sum_{a=1}^4 F(\mneu) 
\eeq
where 
\beq 
F(\mneu) \equiv  {1\over{h'(\mneu)}} \left[ 
  (\mneu)^2 \, c_1  
 + (\mneu) \,  c_2
 +  \mu^2 M_1 M_2 \, c_3 \right] f(m_{\tilde{Z}_a}^2)\,. 
\label{Ffactor}
\eeq
%%%
The analytic expressions for the derivative $ h'(\mneu)$
appearing in the equation above is easily expressed 
in terms of the physical masses and is given by
%%%%%%%%%%%%%%%%
\bea
h'( { m^2_{\tilde{Z}_a} }) \;=\;
\prod_{b \neq a} \; ({ m^2_{\tilde{Z}_a} }\;-\;{ m^2_{\tilde{Z}_b} })
\quad \quad a,b\;=\;1,\ldots ,4  \; \; .
\eea
%%%%%%%%%%%%%%%%%%
The quantities $c_{1,2,3,}$  are given by
\bea
c_1 &=& M_2\,+\,{\tan^2\theta_W} \, M_1  \nonumber \\
c_2 &=& - {\tan^2\theta_W} \, M_1 \, (\mu^2 + M_2^2 ) \,
- \,M_2 \, (\mu^2 + M_1^2 )   \nonumber \\
c_3 &=& M_1\,+\,{\tan^2\theta_W} \, M_2  \, .
\eea
%%%%%%%%%%%%%%%%%%%%%%%%%%%%%%%%%%%%%%%%%%%%%%%%%%%%%%%%%%
%%%%%%%%%%%%%%%%%%%%%%%%%%%%%%%%%%%%%%%%%%%%%%%%%%%%%%%%%% 
 
%\subsection{Gauge and Higgs boson contributions}
For a complete analysis the contributions of the gauge and Higgs bosons to
the quantity $\Delta$, although small, should be also included. Their
contributions can be evaluated more easily using Eq. (\ref{alter}).
In this case the needed
corrections to the pseudo-scalar propagators and tadpoles can be read from
Ref. \cite{Bagger}. 
The pseudo-scalar propagator at zero momentum transfer is given by
%%%%%%%%%%
\bea
16\;\pi^2\;\Pi_{AA}(0)\;&=&\;\frac{g^2}{8}\; \biggl\{ \; 
2\;\tilde{D}(H^+,W)\;+\;\frac{s^2_{\alpha \beta}}{c^2}\;\tilde{D}(H,W)
  \;+\;\frac{c^2_{\alpha \beta}}{c^2}\;\tilde{D}(h,W) \nonumber \\
\;&-&\;
\frac{M_Z^2}{c^2}\; \left[ \; 
c^2_{2 \beta}\;(\; {\bar{c}}^2_{\alpha \beta} \;D(A,H)\;+\;
{\bar{s}}^2_{\alpha \beta} \;D(A,h)\;)\;+\; \right. \nonumber \\
&& \quad \quad \quad
s^2_{2 \beta}\;(\; {\bar{c}}^2_{\alpha \beta} \;D(Z,H)\;+\;
 \left. {\bar{s}}^2_{\alpha \beta} \;D(Z,h)\;)\; \right] \nonumber \\
\;&+&\;
\frac{1}{2 c^2}\;\left[\;c_{2\beta}\;c_{2\alpha}\;(\;f(H)\;-\;f(h)\;)\;-\;
3\;c_{2\beta}^2 \;f(A)\;+\;(1\;-\;3\;s_{2\beta}^2 \;)\;f(Z)\;\right] \nonumber \\
\;&+&\;
\frac{1}{c^2}\;\left[\;(s^2\;c_{2\beta}^2 \;-\;c^2\;(\;1\;+\;s_{2\beta}^2\;))\;f(W)
\;-\;c_{2\beta}^2 \;f(H^+)\;\right]\;  \nonumber \\
\;&-&   \;2\;M_W^2\;D(H^+,W)\;-\;8\;f(W)\;-\;\frac{4}{c^2}\;f(Z)\; \biggr\}  .
\label{paazero}
\eea
%%%%%%%%%%%%
In this expression 
\bea
&&s_{2 \alpha}\;=\;\sin (2 \alpha)\;,\; c_{2 \alpha}\;=\;\cos (2 \alpha)
%\nonumber \\
\;\;,\;\;
s_{2 \beta}\;=\;\sin (2 \beta)\;,\; c_{2 \beta}\;=\;\cos (2 \beta)
\nonumber \\
&&s_{\alpha \beta}\;=\;\sin(\alpha-\beta)\;,\;
c_{\alpha \beta}\;=\;\cos(\alpha-\beta)
%\nonumber \\
\;\;,\;\;
{\bar s}_{\alpha \beta}\;=\;\sin(\alpha+\beta)\;,\;
{\bar c}_{\alpha \beta}\;=\;\cos(\alpha+\beta) \, .   \nonumber 
\eea
The functions appearing in Eq. (\ref{paazero})  are given by
%%%%%%%%
\bea
%%&&f(i)\;=\;2\;m_i^2\;(\; \ln\;( \frac{m_i^2}{Q^2})\;-\;1\;) \;,\;
D(i,j)\;=\;\frac{f(i)\;-\;f(j)}{m_i^2 \;-\;m_j^2}  %%\nonumber \\
\quad ,\quad \tilde{D}(i,j)\;=\;
\frac{m_i^2\;f(i)\;-\;m_j^2\;f(j)}{m_i^2 \;-\;m_j^2}
\eea
%%%
where  $f(i) \equiv f(m_i^2)$. The function $f $
was defined in Eq. (\ref{funf}).
$H$, $h$, $A$, $H^+$ denote the heavy and the light  $CP$-even Higgs, the
pseudo-scalar Higgs and charged Higgs boson respectively.
$W$, $Z$ stand for the charged and neutral gauge bosons respectively.

The combination of the tadpoles needed to calculate
the contributions of the gauge and the Higgs bosons to the quantity
$ \Delta$, through Eq. (\ref{alter}), is given by Ref. \cite{Bagger}
%\newpage
%%%%%%%%%%%%%%%%%
\bea
b_A\;&\equiv& \;\frac{t_1}{v_1} \;s^2_{\beta} \;+\;\frac{t_2}{v_2}
\;c^2_{\beta}
\nonumber \\
\;&=&\; \frac{g^2}{32\;\pi^2}\;\biggl\{
\;-\;\frac{c_{2 \beta}^2}{8 c^2} f(A)\;-\;(\; \frac{1}{2}\;+\;
\frac{c_{2 \beta}^2}{4 c^2} \;)\;f(H^+)\;+\;  \nonumber \\
\;&+&\;\frac{1}{8 c^2}\;
[\;s^2_{\beta}\;(\;c^2_{\alpha} - 3\;s^2_{\alpha} - s_{2 \alpha}\;
\tan \beta\;) \;+\;
c^2_{\beta}\;(\;s^2_{\alpha} - 3\;c^2_{\alpha} - s_{2 \alpha}\;
\cot \beta\;) ]\;f(h)  \nonumber \\
\;&+&\;\frac{1}{8 c^2}\;
[\;s^2_{\beta}\;(\;s^2_{\alpha} - 3\;c^2_{\alpha} + s_{2 \alpha}\;
\tan \beta\;) \;+\;
c^2_{\beta}\;(\;c^2_{\alpha} - 3\;s^2_{\alpha} + s_{2 \alpha}\;
\cot \beta\;) ]\;f(H)  \nonumber \\
\;&+&\;\left(\;\frac{ c^2_{2\beta} }{4 c^2}\;-\;\frac{3}{2}\;\right)\;f(W)
\;+\;\left(\;\frac{ c^2_{2\beta} }{8 c^2}\;
 -\;\frac{3}{4 c^2}\;\right)\;f(Z) \; \biggr\} \,.
\eea
%%%%%%%%%%%%
With this we exhaust the list of all nonvanishing contributions to
the quantity $\Delta$.

For a scale independent result we should
also include the leading logarithmic contributions of the difference
$\;{{\Pi}_{AA}}(0) -  {{\Pi}_{AA}}(m_A) $
involving the renormalization scale $Q$.
%This we discuss in the following.
The $\log(Q^2)$ pieces of this expression, which  
relates the pole and effective potential masses, are listed below. These can
be derived from the analytic expressions found in Ref. \cite{Bagger}.
We write the $\log(Q^2)$ dependent part of this difference as
%%%%%%%%
\bea
\Pi_{Log} \;
&\equiv&\; \biggl[ \; {{{\Pi}_{AA}}(0) -  {{\Pi}_{AA}}(m_A) \;}\biggr]_{\log Q^2}
\nonumber \\ 
\;&=&\; (\;\sum_{i}\; X_{i} \;)\; \frac{m_A^2}{16\; \pi^2}\;
\log\;(\;\frac{m_A^2}{Q^2})  \,.    \label{Log}
\eea
%%%
The remaining part of ${{{\Pi}_{AA}}(0) -  {{\Pi}_{AA}}(m_A)}$
does not involve logarithms of the scale $Q^2$. The nonvanishing
contributions to the quantities $X_{i}$ 
appearing in the expression above are given below
\bea
X_{fermions}\;=\;3 \;c^2_{\beta}\; h^2_t\;+\; 3 \;s^2_{\beta}\; h^2_b\;+\;
\;s^2_{\beta}\; h^2_{\tau} \nonumber\,, \\
X_{charginos}\;=\; g^2 \;,\;X_{neutralinos}\;=\; \frac{g^2}{\cos^2\;\theta_W}
\nonumber\,, \\
X_{gauge + Higgs}\;=\; -\;g^2\;(\;\frac{3}{2}\;+\;\tan^2 \beta\;) \,.
\label{logarithm}
\eea
%%%%%%%
Within the logarithm in the Eq. (\ref{Log}) the pseudo-scalar
mass appears. Any other mass can also lead to a scale independent result.
However we have numerically verified that this choice is the most
appropriate in the sense of better approximating the pole mass.

%%%%%%%%%%%%%%%%%%%%% %%%%%%%%%%%%%%%%%%%%%%%%%
%\newpage
\section{Results and Discussion}
\begin{figure}[t] 
\begin{center} 
\epsfig{file=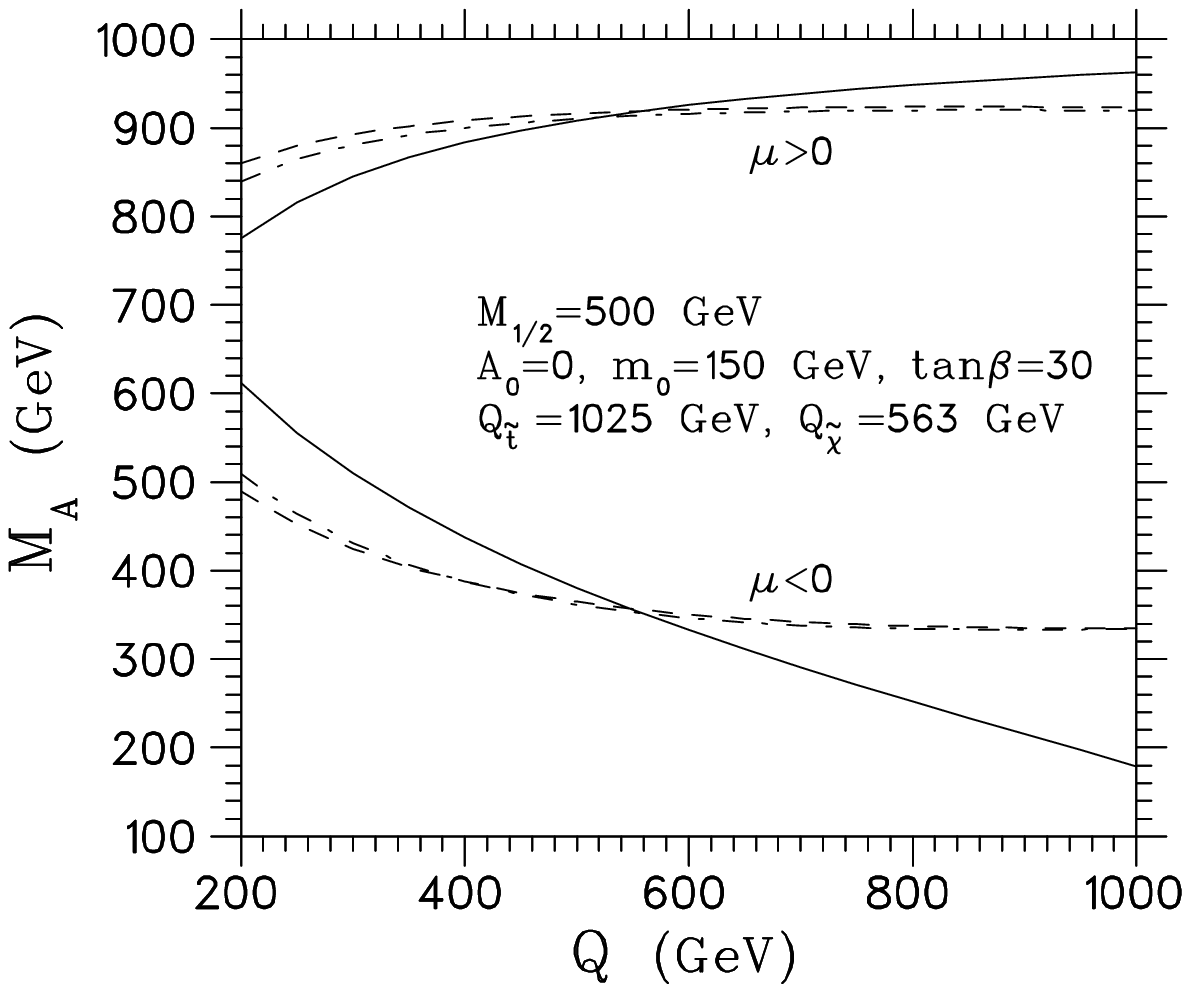,height=9.cm,width=10.cm} 
\vspace{1.cm}
\begin{minipage}[t]{14.cm}  
\caption[]{The pseudo-scalar mass as function of the scale $Q$ for the
inputs displayed in the figure. $Q_{\tilde \chi} (Q_{\tilde t})$ is
the average chargino/neutralino (stop) mass. The solid line is the
effective potential mass where only the third generation of sfermion
contributes. In the dashed line the contribution of all species is taken
into account as well as the the leading log wave function renormalization
contributions. The dashed-dotted line is the pole mass. }
\label{fig1}  
\end{minipage}  
\end{center}  
\end{figure}

Following the previous discussion we see that for defining 
a scale independent pseudo-scalar mass we have to incorporate to the
effective potential mass of Eq. (\ref{mass}) the logarithmic terms given
by Eq. (\ref{Log}). Therefore we define $\; {\tilde{m}}^2_{A} \;$ as
\bea
{\tilde{m}}^2_{A} \; \equiv \;{{m}}^2_{A} \;+\; \Pi_{Log}   \;.
\eea
This is scale independent at the one-loop order and its 
difference from the pole mass is expected to be small.
We have scanned the MSSM parameter space assuming universal
boundary conditions at the Unification scale and have indeed verified that
$ {\tilde{m}}_{A} $ is very close to the pole mass. Differences are
less than $2 \%$.
We have also seen that the logarithmic corrections  $\Pi_{Log}$ 
are small and therefore not very significant. Thus  
the bulk of the radiative corrections to ${\tilde{m}}_{A} $
is carried by the effective potential mass $m_A$. 
A sample result depicting
very clearly the situation is displayed in the Figure \ref{fig1}. 
For the given inputs we plot the mass  ${\tilde{m}}_{A} $
as function of the scale $Q$ (dashed line) and compare it to  
the one-loop pole mass (dashed-dotted line). 
${\tilde{m}}_{A}$ is indeed very close to the pole mass and almost
independent of the scale $Q$ for values of $Q \;\geq\;Q_{\tilde \chi}$.
In the displayed figures 
the solid line corresponds to the value of the effective potential mass 
$m_A$ where the contribution of charginos and neutralinos is omitted. 
For definiteness we shall denote this by $m_A(0)$. This 
is scale dependent and deviates from the pole mass. This deviation becomes 
significant for large values of $\; M_{1/2} \gg  m_0\;$ and large 
$\tan \beta$.
In some cases the deviation of
${{m}}_{A}(0) $  from the pole mass can be as large as $25 \% $
depending on the scale it is calculated. This shows the significance
of the chargino and neutralino corrections in estimating the
pseudo-scalar Higgs boson mass. 
Only at $Q \approx Q_{\tilde \chi} $ this coincides with the pole mass.

Figure 2 shows the same situation for the same values of the soft masses
and trilinear couplings but for a smaller value of $\tan \beta$.
The effect is less dramatic and the inclusion of the charginos and
neutralinos, denoted by $\tilde {\chi }$ in the sequel, amounts to no more
than $ \approx 3 \%$. In Figure 3 we have taken $M_{1/2}$ to be comparable
to $m_0$ but the angle $\tan \beta$ is taken to be large ($=30$).
In this case
the resulting $\tilde {\chi}$ corrections are of the order of
$\approx 7 \%$. In all figures the mass $m_A(0)$, unlike
$\;{\tilde{m}}_{A}$, is scale dependent resulting to a poor determination
of the pseudo-scalar mass.
From the discussion above it becomes obvious that the effect is more
pronounced in regions of the parameter space for which both $ M_{1/2} $ and
$\tan \beta$ are large. The reason is that in the large  $\tan \beta$
regime $m_A^2$ is given by
%%%
\bea
m_A^2\;\approx \;- \tan \beta \;(m_3^2\;+\;\Delta) \,.  \; \; \label{qq1}
\eea
%%%%%%%%%%%%%%%%%
Therefore the gaugino soft terms of Eq. (\ref{rge}) that enter the RGE
of the parameter $m_3^2$, 
which become important for large values of the
parameter $M_{1/2}$, give a large scale dependence if not cancelled by
corresponding terms in the corrections $\Delta$. Failing to
include the $\tilde \chi$ contributions results to incomplete cancellations
of $\log (Q^2)$ terms in (\ref{qq1}), and this effect is augmented for
large values of $\tan \beta$. This is the reason the mass $m_A(0)$ shows
this drastic scale dependence.

The masses ${\tilde{m}}_{A} $  and ${{m}}_{A}(0)  $ are almost
equal at scales comparable to $Q_{\tilde {\chi}}$ and thus both very close
to the pole mass. This is expected since at $Q \approx Q_{\tilde {\chi}}$
the contribution of charginos and neutralinos are small and the
$ \Pi_{Log}$ terms, as said before, 
do not contribute significantly. The smallness of the
$\tilde {\chi }$'s contributions at $Q \approx Q_{\tilde {\chi}}$
is due to the fact that their masses are
close to $Q_{\tilde {\chi}}$  resulting to small logarithmic corrections
within the expression of the effective potential. Therefore neglecting the
contribution of the charginos and neutralinos is legitimate provided that
the relevant scale of the calculation is taken to be
$Q_{\tilde {\chi}}$ rather than $Q_{\tilde {t}}$.
These two scales however may lie quite apart.
In the displayed figures the equality of the two masses
is at the point where the dashed and the dashed curve cross. One
sees that this scale is quite close to $Q_{\tilde {\chi}}$ whose value is
shown in the figures.

Below  $Q_{\tilde {\chi}}$  and $Q_{\tilde {t}}$, and as we approach
$M_Z$, the mass ${\tilde{m}}_{A} $ shows an unstable behaviour with
changing the scale $Q$ which is however milder than that of $m_A(0)$. This
was expected since for such values of $Q$, which lie lower than the
$\tilde{t}$ and ${\tilde {\chi}}$ masses, large logarithms enter the
one-loop corrections of the effective potential. At such scales 
$Q$ only the inclusion of the two-loop corrections can ameliorate
the situation and yield a stable result. 

We conclude that for an unambiguous determination of the pseudo-scalar
mass, as this is calculated from the effective potential,
the contribution of the charginos and neutralinos
should be taken into account. Their contributions is 
significant in regions of the parameter space for which $M_{1/2}$ and
$\tan \beta$ are large. Neglecting their contributions can be legitimate
only if the calculation is performed at a scale close to the average of
the chargino and neutralino masses. Adding the small contributions of the
remaining sectors, gauge and Higgs bosons, as well as, 
the small leading logarithmic corrections 
of the difference $\Pi(0) -\Pi(m_A^2)$, results to a mass
which is scale independent and approximates accurately the pole mass.

\vspace{.9cm}
\noindent 
{\bf Acknowledgments} \\ 
\noindent 
A.B.L. acknowledges support from ERBFMRXCT--960090 TMR programme
and D.V.N. by D.O.E. grant DE-FG03-95-ER-40917.
V.C.S. acknowledges support  from Academy of Athens under 200/486-2000 grant.
A.B.L.  wishes to express his thanks to A. Pilaftsis for communications.

%%%%%%%%%%%%%%%%%%%%%%%
\clearpage
%%%%%%%%%%%%%%%%%%%%%%%%%% Figure 1 %%%%%%%%%%%%%%%%%%%%%%%%%%%%%%%% 

\clearpage
%%%%%%%%%%%%%%%%%%%%%%%%%% Figure 2 %%%%%%%%%%%%%%%%%%%%%%%%%%%%%%%% 
\begin{figure}[t] 
\begin{center} 
\epsfig{file=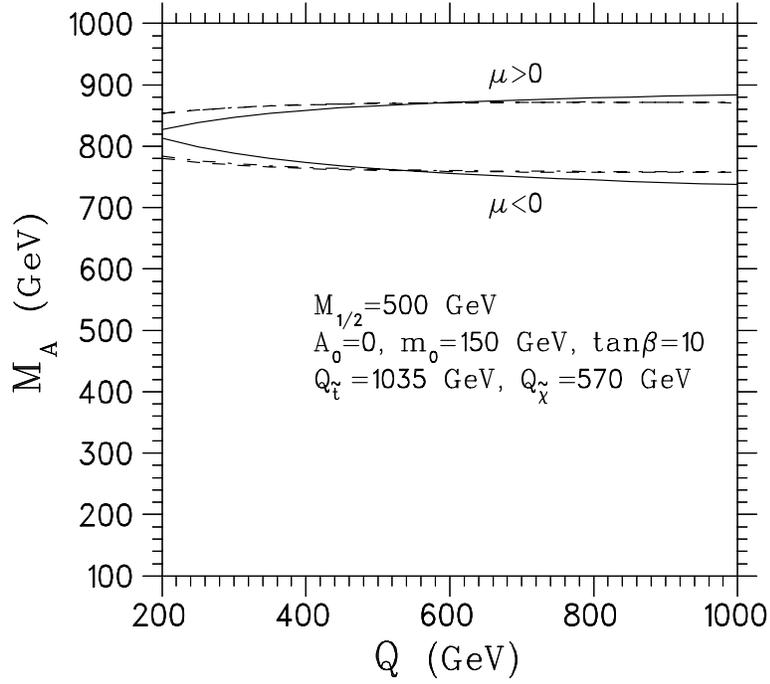,height=9.cm,width=10.cm}
 
\vspace{1.cm}
\begin{minipage}[t]{14.cm}  
\caption[]{ The same as in Figure \ref{fig1} with $\tan \beta=10$.}
\label{fig2}  
\end{minipage}  
\end{center}  
\end{figure}

%%%%%%%%%%%%%%%%%%%%%%%%%% Figure 3 %%%%%%%%%%%%%%%%%%%%%%%%%%%%%%%% 
\begin{figure}[t] 
\begin{center} 
\epsfig{file=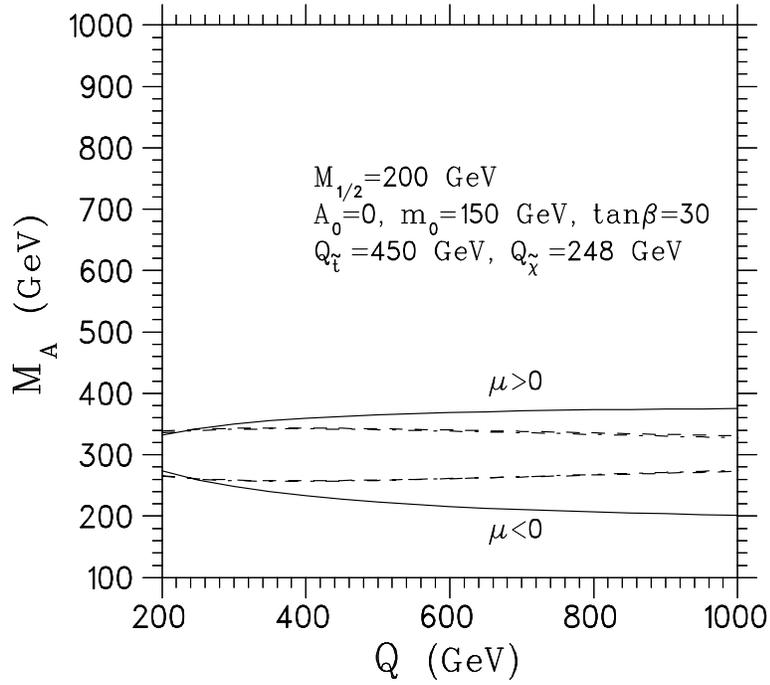,height=9.cm,width=10.cm}
 
\vspace{1.cm}
\begin{minipage}[t]{14.cm}  
\caption[]{ The same an in Figure \ref{fig1} with $M_{1/2}$ to be
comparable to $m_0$.}
\label{fig3}  
\end{minipage}  
\end{center}  
\end{figure}

\end{document}